\documentclass[12pt]{article}
\usepackage{a4}
\usepackage{amssymb}
\usepackage{cite}
\newcommand{\be}{\begin{equation}}
\newcommand{\ee}{\end{equation}}
\newcommand{\bea}{\begin{eqnarray}}
\newcommand{\eea}{\end{eqnarray}}

\usepackage{hyperref}
\begin{document}
\thispagestyle{empty}
\def\thefootnote{\fnsymbol{footnote}}
\begin{center}\Large
Topological defects
in the Liouville field theories with different cosmological constants. \\
\vskip 2em
February 2018
\end{center}\vskip 0.2cm
\begin{center}
Elena Apresyan$^{1}$
\footnote{elena-apresyan@mail.ru}
 and Gor Sarkissian$^{2,3}$\footnote{ gor.sarkissian@ysu.am, sarkissn@theor.jinr.ru}
\end{center}
\vskip 0.2cm

\begin{center}
$^1$ Yerevan Physics Institute, \\
Alikhanian Br. 2, 0036\, Yerevan\\
Armenia
\end{center}
\begin{center}
$^2$Bogoliubov Laboratory of Theoretical Physics, JINR\\
141980 Dubna, Moscow region\\
Russia\\
\end{center}
\begin{center}
$^3$Department of Physics, \ Yerevan State University,\\
Alex Manoogian 1, 0025\, Yerevan\\
Armenia
\end{center}

\vskip 1.5em
\begin{abstract} \noindent
We construct topological defects in the Liouville field theory producing
jump in the value of cosmological constant. We construct them using the Cardy-Lewellen
equation for the two-point function with defect. We show that there are
continuous and discrete families of such kind of defects.
For the continuous family of defects  we also find the Lagrangian description
and check its agreement with the  solution of the Cardy-Lewellen equation using the heavy asymptotic
semiclasscial limit.
 \end{abstract}
\newpage
\newpage
\section{Introduction}
In this paper we construct topological defects gluing 2D Liouville field theories with different cosmological constants.

Topological defects in the Louville field theory with the same cosmological constants on the both side were constructed by the second author
almost ten years back in papers \cite{Sarkissian:2009aa,Sarkissian:2011tr}. In that papers two-point functions in the presence of defects
were computed using the Cardy-Lewellen equation for defects. It was derived that there exist two families of defects, discrete, with one-dimensional world-volume,
and continuous, with two-dimensional world-volume. Later for the continuous family of defects also the Lagrangian description was suggested in \cite{Aguirre:2013zfa}. It was shown in \cite{Poghosyan:2015oua} that this Lagrangian description  is in agreement with the
found in \cite{Sarkissian:2009aa,Sarkissian:2011tr} defect two-point function using various semiclassical limits.

Here we generalize above mentioned calculations to the case of the different cosmological constants.
First we elaborate to this case the Cardy-Lewellen relation for defects. We find that in fact the two-point functions are given
by the same functions as before but which get rescaled by the factor $\left(\mu_2\over \mu_1\right)^{-iP\over b}$, where
$\mu_1$ and $\mu_2$ are the cosmological constants, and $P$ is a momentum.
Formulae (\ref{discret}) and (\ref{cont}) are our main results.
For the continuous family of defect we also constructed the corresponding Lagrangian and checked via the heavy asymptotic limit
that it is in agreement with the two-point function (\ref{cont}).

We would like to say that one of the motivations of this research was recently suggested in papers  \cite{Vafa:2015euh,Can:2014tca}
the idea to describe the Fractional Quantum Hall effect (FQHE) by the Liuoville field theory, whose cosmological constant should play a role
of a chemical potential. On the other hand it is known that in the FQHE one has jump of the chemical potential \cite{girvin}.
We have an impression that our construction which in fact connects two Liouville theories with the different cosmological constants
may have an application to the FQHE.

The paper is organized as follows. In section 2 we collect several necessary for us facts on classical and quantum Liouville field theory.
In section 3 we generalized and solved  the Cardy-Lewellen equation for defects to the case of the different cosmological constants.
We showed also here that constructed defects indeed map FZZ \cite{Fateev:2000ik} and ZZ \cite{Zamolodchikov:2001ah} boundary states of the first theory to the linear combinations of the FZZ and ZZ boundary states of the second theory.
In section 4 we have written down the Lagrangian for the continuous family of defects. In section 5 we checked, using the heavy asymptotic
semiclassical limit, that the two-point function for the continuous family of defects computed in section 3, is in agreement with the Lagrangian description of section 4.

\section{Review of Liouville theory}

Let us recall some basic facts on classical and quantum  Liouville theory.

The action of the Liouville theory is

\be\label{azione}
S={1\over 2\pi i}\int\left(\partial\phi\bar{\partial} \phi+\mu \pi e^{2b\phi}\right)d^2 z\, .
\ee
Here we use a complex coordinate $z=\tau+i\sigma$, and $d^2z\equiv dz\wedge d\bar{z}$ is the volume form.

The field $\phi(z,\bar{z})$ satisfies the  Liouville equation:
\be\label{leom}
\partial\bar{\partial}\phi=\pi\mu b e^{2b\phi}\, .
\ee
The general solution to  (\ref{leom}) can be written in terms
of two arbitrary functions $A(z)$ and $B(\bar{z})$:
\be\label{liusol}
\phi={1\over 2b}\log\left({1\over \pi\mu b^2}{\partial A(z)\bar{\partial}B(\bar{z})
\over (A(z)+B(\bar{z}))^2}\right)\, .
\ee
The solution (\ref{liusol}) is invariant if one transforms $A$ and $B$
simultaneously by the  constant M\"{o}bius transformations:
\be\label{mobsim}
A\to{\zeta A+\beta\over \gamma A+\delta},\hspace{0.5cm}
B\to{\zeta B-\beta\over -\gamma B+\delta},\hspace{0.5cm}
\zeta\delta-\beta\gamma=1\, .
\ee
Classical expressions for the holomorphic and anti-holomorphic components of the energy-momentum tensor are
\be\label{tl}
T=-(\partial\phi)^2+b^{-1}\partial^2\phi\, ,
\ee
\be\label{tr}
\bar{T}=-(\bar{\partial}\phi)^2+b^{-1}\bar{\partial}^2\phi\, .
\ee
Inserting (\ref{liusol}) in (\ref{tl}) and (\ref{tr})
we obtain, that components of the energy-momentum tensor are
 given  by the Schwarzian derivatives of $A(z)$ and $B(\bar{z})$:
\be
T=\{A; z\}={1\over 2b^2}\left[{A'''\over A'}-{3\over 2}{(A'')^2\over (A')^2}\right]\, ,
\ee
\be
\bar{T}=\{B; \bar{z}\}={1\over 2b^2}\left[{B'''\over B'}-{3\over 2}{(B'')^2\over (B')^2}\right]\, .
\ee
The Schwarzian derivative is invariant under a constant
 M\"{o}bius transformation:
\be\label{trmo}
\left\{{\zeta F+\beta\over \gamma F+\delta}; z\right\}=\{F;z\}
,\hspace{0.5cm}
\zeta\delta-\beta\gamma=1\, .
\ee

Quantum Liouville  field theory is a conformal field theory enjoying the Virasoro algebra
\be
[L_m,L_n]=(m-n)L_{m+n}+{c_L\over 12}(n^3-n)\delta_{n,-m}\, ,
\ee
with the central charge
\be
c_L=1+6Q^2\, .
\ee

Two-point functions of Liouville theory are given by the  function $S(\alpha)$
:
\be\label{twopf}
\langle V_{\alpha}(z_1,\bar{z}_1)V_{\alpha}(z_2,\bar{z}_2)\rangle={S(\alpha)\over (z_1-z_2)^{2\Delta_{\alpha}}
(\bar{z}_1-\bar{z}_2)^{2\Delta_{\alpha}}}\, ,
\ee
\be\label{reflal}
S(\alpha)={\left(\pi\mu\gamma(b^2)\right)^{b^{-1}(Q-2\alpha)}\over b^2}{\Gamma(1-b(Q-2\alpha))\Gamma(-b^{-1}(Q-2\alpha))
\over \Gamma(b(Q-2\alpha))\Gamma(1+b^{-1}(Q-2\alpha))}\, .
\ee
The spectrum of the Liouville theory has
the  form
\be\label{lsdi}
{\cal H}=\int_0^{\infty} dP \;R_{{Q\over 2}+iP}\otimes R_{{Q\over 2}+iP}\, ,
\ee
where $R_{\alpha}$ is the highest weight representation with respect to the Virasoro algebra.

\section{Two-point function with defect producing jump in cosmological constant}

As we mentioned before the aim of this work is to construct topological defect gluing
two Liouville theories with different cosmological constants. For this purpose
we will use bootstrap programm. In fact the bootstrap programm for topological defects with the same theory on both sides
was developed in \cite{Petkova:2001ag,Sarkissian:2009aa,Sarkissian:2011tr}. Here we will reconsider it, taking into account the necessary changes caused by the presence of two different
cosmological constants on the different sides of the defects.

We consider a topological defect mapping the Hilbert space of the first theory on the Hilbert space of the second theory $X: H_{(1)}\to H_{(2)}$
in the form:
\be
X=\int_{{Q\over 2}+i\mathbb{R}} d\alpha\,{\cal D}(\alpha)\,\mathbb{P}^{\alpha}\, ,
\ee
where $\mathbb{P}^{\alpha}$ are maps:
\be
\mathbb{P}^{\alpha}=\sum_{N,M}(|\alpha,N\rangle_{(2)}\otimes \overline{|\alpha,M\rangle}_{(2)})
(_{(1)}\langle \alpha,N|\otimes_{(1)} \overline{\langle\alpha,M|})\, .
\ee
Here $|\alpha,N\rangle_{(i)}$ and $\overline{|\alpha,M\rangle}_{(i)}$, $i=1,2$ are vectors of  orthonormal bases of left and right copy of $R_{\alpha}$
 of the first and second theory respectively.
Two-point functions with a defect $X$ inserted can be written as
\be\label{twopff}
\langle V^{(2)}_{\alpha}(z_1,\bar{z}_1)X V^{(1)}_{\alpha}(z_2,\bar{z}_2)\rangle={D^{\alpha}\over
(z_1-z_2)^{2\Delta_{\alpha}}(\bar{z}_1-\bar{z}_2)^{2\Delta_{\alpha}}}\, .
\ee
where
\be
D^{\alpha}={\cal D}(\alpha)S^{(1)}(\alpha)
\ee

Consider the following four-point function with the defects insertions:
\be\label{fourp}
\langle V^{(2)}_{-b/2}(z_1,\bar{z}_1)V^{(2)}_{\alpha}(z_2,\bar{z}_2)XV^{(1)}_{\alpha}(z_3,\bar{z}_3)V^{(1)}_{-b/2}(z_4,\bar{z}_4)X^{\dagger}\rangle\, .
\ee
One can compute (\ref{fourp}) in two pictures.
In the first picture at the beginning we use the OPE
\be\label{ope}
V^{(j)}_{\alpha}V^{(j)}_{-b/2}\sim C_{-b/2,\alpha}^{(j)\,\alpha-b/2}V^{(j)}_{\alpha-b/2}+C_{-b/2,\alpha}^{(j)\,\alpha+b/2}V^{(j)}_{\alpha+b/2}\,.\,\quad
j=1,2
\ee
 and then (\ref{twopff}) for the fields produced in this process.
This results in

\be\label{frprd}
\sum_{\pm} D^{\alpha\pm b/2}D^0
C_{-b/2,\alpha}^{(1)\,\alpha\pm b/2}C_{-b/2,\alpha}^{(2)\,\alpha\pm b/2}\left({\cal F}_{\alpha\pm b/2}\left[\begin{array}{cc}
\alpha &\alpha\\
-b/2&-b/2 \end{array}\right]\right)^2\, ,
\ee
where ${\cal F}_{\alpha\pm b/2}\left[\begin{array}{cc}
\alpha & \alpha \\
-b/2&-b/2 \end{array}\right]$ is so called conformal block  giving contribution of
the descendant fields in the OPE (\ref{ope}). It appears squared since it is separately produced by the left and right modes.

In the second picture we move the field $V^{(2)}_{-b/2}(z_1,\bar{z}_1)$ to the most right position:
\bea\label{secprd}
\langle V^{(2)}_{\alpha}(z_2,\bar{z}_2)X V^{(1)}_{\alpha}(z_3,\bar{z}_3)V^{(1)}_{-b/2}(z_4,\bar{z}_4)X^{\dagger}V^{(2)}_{-b/2}(z_1,\bar{z}_1)\rangle
\eea
and then use twice (\ref{twopff}) resulting in
\be
D^{\alpha}D^{-b/2}
\left({\cal F}_{0}\left[\begin{array}{cc}
\alpha&-b/2\\
\alpha&-b/2 \end{array}\right]\right)^2
+\cdots\, .
\ee

Using the fusing matrix:
\bea
{\cal F}_{\alpha\pm b/2}\left[\begin{array}{cc}
\alpha&\alpha\\
-b/2&-b/2 \end{array}\right]
=
 F_{\alpha\pm b/2 0}\left[\begin{array}{cc}
-b/2&-b/2\\
\alpha&\alpha\end{array}\right] {\cal F}_{0}\left[\begin{array}{cc}
\alpha&-b/2\\
\alpha&-b/2 \end{array}\right]+\cdots\, ,
\eea
we obtain
\bea\label{defalg}
\sum_{\pm} D^0D^{\alpha\pm b/2}C_{-b/2,\alpha}^{\alpha\pm b/2(1)}C_{-b/2,\alpha}^{\alpha\pm b/2(2)}\left(
 F_{\alpha\pm b/20}\left[\begin{array}{cc}
-b/2&-b/2\\
\alpha&\alpha\end{array}\right]\right)^2=
D^{\alpha}D^{-b/2}\, .\hspace{1cm}
\eea
This is the Cardy-Lewellen cluster condition for defects.

Let us use the relation \cite{Sarkissian:2011tr}:
\be\label{cfal}
C_{\alpha_1,\alpha_2}^{(j)\, \alpha_3}F_{\alpha_3,0}\left[\begin{array}{cc}
\alpha_1&\alpha_1\\
\alpha_2 &\alpha_2 \end{array}\right]=W^{(j)}(0){W^{(j)}(\alpha_3)\over W^{(j)}(\alpha_1)W^{(j)}(\alpha_2)}\, ,\quad j=1,2
\ee
where $W^{(j)}(\alpha)$ is the ZZ function \cite{Zamolodchikov:2001ah}:
\be\label{zzfu}
W^{(j)}(\alpha)=-{2^{3/4}e^{3i\pi/2}(\pi\mu_j\gamma(b^2))^{-{(Q-2\alpha)\over 2b}}\pi(Q-2\alpha)\over
\Gamma(1-b(Q-2\alpha))\Gamma(1-b^{-1}(Q-2\alpha))}\, ,\quad j=1,2
\ee
Define $\Psi(\alpha)$ by the equation:
\be\label{dkdoll}
{D^{\alpha}\over D^0}=\Psi(\alpha){W^{(1)}(0)W^{(2)}(0)\over W^{(1)}(\alpha)W^{(2)}(\alpha)}\, .
\ee
For $\Psi(\alpha)$ the equation (\ref{defalg}) takes the form:
\be\label{psik}
\Psi(\alpha)\Psi(-b/2)=\Psi(\alpha-b/2)+\Psi(\alpha+b/2)\, ,
\ee
The solution of the equation (\ref{psik}) is
\be
\Psi_{m,n}(\alpha)={\sin(\pi m b^{-1}(2\alpha- Q))\sin(\pi nb (2\alpha-Q))\over \sin(\pi mb^{-1} Q)\sin(\pi nb Q)}\, ,
\ee
Using (\ref{dkdoll}) we obtain for the defect two-point function:

\be\label{discrett}
D_{m,n}(\alpha)={\sin(\pi m b^{-1}(2\alpha- Q))\sin(\pi nb (2\alpha-Q))\over W^{(1)}(\alpha)W^{(2)}(\alpha)}\, .
\ee
And finally dividing on $S^{(1)}(\alpha)$:
\be\label{reflal}
S^{(1)}(\alpha)={\left(\pi\mu_1\gamma(b^2)\right)^{b^{-1}(Q-2\alpha)}\over b^2}{\Gamma(1-b(Q-2\alpha))\Gamma(-b^{-1}(Q-2\alpha))
\over \Gamma(b(Q-2\alpha))\Gamma(1+b^{-1}(Q-2\alpha))}\, .
\ee

 we get
\be\label{discret}
{\cal D}_{m,n}(\alpha)=\left(\mu_2\over \mu_1\right)^{Q-2\alpha\over 2b}{\sin(\pi m b^{-1}(2\alpha- Q))\sin(\pi nb (2\alpha-Q))\over \sin\pi b^{-1}(2\alpha-Q)\sin\pi b(2\alpha-Q)}\, .
\ee

But this is not the end of the story.

Assume that we have a family of defects parameterized by $\kappa$.
In this case  $D^{-b/2}/ D^0$, which is
two-point function of the degenerate field $V_{-b/2}$ in the presence of defect, will be a function $A(\kappa,b)$
of $\kappa$ and $b$. Substituting
\be
A={D^{-b/2}\over D^0}
\ee
and
\be\label{dkdol}
D^{\alpha}={\Lambda^{\alpha}\over W^{(1)}(\alpha)W^{(2)}(\alpha)}\, .
\ee
in (\ref{psik}) we obtain a linear equation for $\Lambda(\alpha)$:

\be\label{tsik}
{W^{(1)}(-b/2)W^{(2)}(-b/2)\over W^{(1)}(0)W^{(2)}(0)} A\Lambda(\alpha)=\Lambda(\alpha-b/2)+\Lambda(\alpha+b/2)
\ee

The solution of (\ref{tsik}) is indeed one-parametric family,
\be\label{funcla}
\Lambda_s(\alpha)=\cosh(2\pi s(2\alpha-Q))\, ,
\ee
 with parameter $s$ related to $A$ by
\be\label{rela}
2\cosh 2\pi bs=A{W^{(1)}(-b/2)W^{(2)}(-b/2)\over W^{(1)}(0)W^{(2)}(0)}\, .
\ee

Substituting (\ref{funcla}) in (\ref{dkdol}) we obtain for $ D_{s}(\alpha)$ and ${\cal D}_{s}(\alpha)$ respectively
\be\label{contt}
D_{s}(\alpha)=-{2^{1/2}i\cosh(2\pi s(2\alpha-Q))\over W^{(1)}(\alpha) W^{(2)}(\alpha)}\, .
\ee
\be\label{cont}
{\cal D}_{s}(\alpha)=\left(\mu_2\over \mu_1\right)^{Q-2\alpha\over 2b}{\cosh(2\pi s(2\alpha-Q))\over 2\sin\pi b^{-1}(2\alpha-Q)\sin\pi b(2\alpha-Q)}\, .
\ee
So we have two groups of topological defects:
\be\label{mu1}
X_s=\int_{{Q\over 2}+i\mathbb{R}} d\alpha\,{\cal D}_s(\alpha)\,\mathbb{P}^{\alpha}\, ,
\ee
\be\label{mu2}
X_{m,n}=\int_{{Q\over 2}+i\mathbb{R}} d\alpha\,{\cal D}_{m,n}(\alpha)\,\mathbb{P}^{\alpha}\, ,
\ee
Recall that in each copy of the  Liouville field theory one has two groups of boundary states, FZZ states \cite{Fateev:2000ik}:
\be
|s\rangle^{(j)}=\int_{{Q\over 2}+i\mathbb{R}} B_{ s}^{(j)}(\alpha)|\alpha\rangle\rangle_{(j)} d\alpha
\ee
and ZZ states \cite{Zamolodchikov:2001ah}:
\be
|m,n\rangle^{(j)}=\int_{{Q\over 2}+i\mathbb{R}} B_{m,n}^{(j)}(\alpha)|\alpha\rangle\rangle_{(j)} d\alpha
\ee
where $|\alpha\rangle\rangle_{(j)}$ are the Ishibashi states satisfying  $L_n^{(j)}+\bar{L}_{-n}^{(j)}=0$, and
\be\label{conttt}
B^{(j)}_{s}(\alpha)=-{2^{1/2}i\cosh(2\pi s(2\alpha-Q))\over W^{(j)}(\alpha)}\, .\quad j=1,2
\ee
\be\label{discrett}
B^{(j)}_{m,n}(\alpha)={\sin(\pi m b^{-1}(2\alpha- Q))\sin(\pi nb (2\alpha-Q))\over W^{(j)}(\alpha)}\, .\quad j=1,2
\ee

Using the identities:

\be\label{iden1}
\sinh(2\pi nbP)\sinh(2\pi n'bP)=\sum_{l=0}^{{\rm min}(n,n')-1}\sinh(2\pi bP)\sinh(2\pi b (n+n'-2l-1)P)
\ee
and
\be\label{iden2}
{\sinh(2\pi nbP)\over \sinh(2\pi bP)}=\sum_{l=1-n,2}^{n-1}\exp(2\pi lbP)
\ee
 one obtains that fusion of the defects (\ref{mu1}), (\ref{mu2}) with boundary state of the first theory producing linear combination
 of the boundary states of the second theory:
\be
X_{m,n}|m',n'\rangle^{(1)} =\sum_{l=0}^{{\rm min}(n,n')-1}
\sum_{k=0}^{{\rm min}(m,m')-1}|m+m'-2k-1,n+n'-2l-1\rangle^{(2)}
\ee

\be
X_{m,n}|s\rangle^{(1)} =\sum_{l=1-n,2}^{n-1}\sum_{k=1-m,2}^{m-1}|s+i(k/b+lb)/2\rangle^{(2)}
\ee

\be
X_{s}|m,n\rangle^{(1)}=\sum_{l=1-n,2}^{n-1}\sum_{k=1-m,2}^{m-1}|s+i(k/b+lb)/2\rangle^{(2)}
\ee
as it is indeed expected.

\section{Lagrangian of the Liouville theory with defect $X_s$}
 We propose the following  action for the Liouville theories
 with the different cosmological constants connected by the topological defect:
\bea\label{topdef}
&&S^{\rm top-def}={1\over 2\pi i}\int_{\Sigma_1}\left(\partial\phi_1\bar{\partial} \phi_1+\mu_1 \pi e^{2b\phi_1}\right)d^2 z+
{1\over 2\pi i}\int_{\Sigma_2}\left(\partial\phi_2\bar{\partial} \phi_2+\mu_2 \pi e^{2b\phi_2}\right)d^2 z\quad\quad\quad\\ \nonumber
&&+ \int_{\partial \Sigma_1}\left[-{1\over 2\pi}\phi_2\partial_{\tau}\phi_1+
{1\over 2\pi}\Lambda\partial_{\tau}(\phi_1-\phi_2)+{\sqrt{\mu_1\mu_2}\over 2}
e^{(\phi_1+\phi_2-\Lambda)b}\right.\\ \nonumber
&&-{1\over \pi b^2}
e^{\Lambda b}\left.\left({1\over 2}\sqrt{\mu_1\over \mu_2}e^{(\phi_1-\phi_2)b}+{1\over 2}\sqrt{\mu_2\over \mu_1}e^{-(\phi_1-\phi_2)b}-\kappa\right)\right]{d\tau\over i}\, .
\eea
Here $\Sigma_1$ is the upper half-plane $\sigma={\rm Im} z\geq 0$ and
$\Sigma_2$ is the lower  half-plane $\sigma={\rm Im} z\leq 0$.
The defect is located along their common boundary,
which is the real axis $\sigma=0$ parameterized by $\tau={\rm Re} z$.
Note that $\Lambda(\tau)$ here is an additional
field associated with the defect itself.
In fact this is the Lagrangian proposed in \cite{Aguirre:2013zfa} and considered in detail in \cite{Poghosyan:2015oua}, but which is modified in a way to take into account
that now $\mu_1\neq \mu_2$, and which becomes the old one for $\mu_1= \mu_2$.

The action (\ref{topdef}) yields the following defect equations of motion at $\sigma=0$:
\bea\label{eom1}
&&{1\over 2\pi}(\partial-\bar{\partial})\phi_1+
{1\over 2\pi}\partial_{\tau}\phi_2-
{1\over 2\pi}\partial_{\tau}\Lambda+
{\sqrt{\mu_1\mu_2} b\over 2}e^{(\phi_1+\phi_2-\Lambda)b}\\ \nonumber
&&-{1\over \pi b}e^{\Lambda b}\left({1\over 2}\sqrt{\mu_1\over \mu_2}e^{(\phi_1-\phi_2)b}-{1\over 2}\sqrt{\mu_2\over \mu_1}e^{-(\phi_1-\phi_2)b}\right)=0\, ,
\eea

\bea\label{eom2}
&&-{1\over 2\pi}(\partial-\bar{\partial})\phi_2-
{1\over 2\pi}\partial_{\tau}\phi_1+
{1\over 2\pi}\partial_{\tau}\Lambda+
{\sqrt{\mu_1\mu_2} b\over 2}e^{(\phi_1+\phi_2-\Lambda)b}\\ \nonumber
&&+{1\over \pi b}e^{\Lambda b}\left({1\over 2}\sqrt{\mu_1\over \mu_2}e^{(\phi_1-\phi_2)b}-{1\over 2}\sqrt{\mu_2\over \mu_1}e^{-(\phi_1-\phi_2)b}\right)=0\, ,
\eea

\be\label{eom3}
{1\over 2\pi}\partial_{\tau}(\phi_1-\phi_2)-{\sqrt{\mu_1\mu_2} b\over 2}
e^{(\phi_1+\phi_2-\Lambda)b}-{1\over \pi b}
e^{\Lambda b}\left({1\over 2}\sqrt{\mu_1\over \mu_2}e^{(\phi_1-\phi_2)b}+{1\over 2}\sqrt{\mu_2\over \mu_1}e^{-(\phi_1-\phi_2)b}-\kappa\right)=0\, .
\ee
The last equation is derived calculating variation by the $\Lambda$.

Using that $\partial_{\tau}=\partial+\bar{\partial}$
and  forming various linear combinations of  equations (\ref{eom1})-(\ref{eom3})
we can bring them
to the form:
\be\label{def3}
\bar{\partial}(\phi_1-\phi_2)=
\pi\sqrt{\mu_1\mu_2} b e^{b(\phi_1+\phi_2)}e^{-\Lambda b}\, ,
\ee
\be\label{kapik}
\partial(\phi_1-\phi_2)={2\over b}e^{\Lambda b}
\left({1\over 2}\sqrt{\mu_1\over \mu_2}e^{(\phi_1-\phi_2)b}+{1\over 2}\sqrt{\mu_2\over \mu_1}e^{-(\phi_1-\phi_2)b}-\kappa\right)\, .
\ee
\be\label{def11}
\partial(\phi_1+\phi_2)-\partial_{\tau}\Lambda=
{2\over b}e^{\Lambda b}\left({1\over 2}\sqrt{\mu_1\over \mu_2}e^{(\phi_1-\phi_2)b}-{1\over 2}\sqrt{\mu_2\over \mu_1}e^{-(\phi_1-\phi_2)b}\right)\, .
\ee

Let us require, that $\Lambda$ is restriction to the real axis of a holomorphic
field
\be\label{hollam}
\bar{\partial}\Lambda=0\, .
\ee
This condition allows to rewrite (\ref{def11}) in the form
\be\label{def111}
\partial(\phi_1+\phi_2-\Lambda)=
{2\over b}e^{\Lambda b}\left({1\over 2}\sqrt{\mu_1\over \mu_2}e^{(\phi_1-\phi_2)b}-{1\over 2}\sqrt{\mu_2\over \mu_1}e^{-(\phi_1-\phi_2)b}\right)\, .
\ee
We can check
that the system of the defect equations of motion (\ref{def3})-(\ref{def111})
guarantees that both components of the energy-momentum tensor
are continuous across the defects and therefore
describes
topological defects:
 \be\label{tl1}
-(\partial\phi_1)^2+b^{-1}\partial^2\phi_1=
-(\partial\phi_2)^2+b^{-1}\partial^2\phi_2\, ,
\ee
\be\label{tr2}
-(\bar{\partial}\phi_1)^2+b^{-1}\bar{\partial}^2\phi_1=
-(\bar{\partial}\phi_2)^2+b^{-1}\bar{\partial}^2\phi_2\, .
\ee
 Therefore, remembering that the solution  (\ref{liusol}) is invariant under the transformation (\ref{mobsim}), and that
  the chiral components of the energy-momentum tensor are invariant under the M\"obius transformation (\ref{trmo}), we can without loosing generality look for
a solution in the form:
\be\label{phi1}
\phi_1={1\over 2b}\log\left({1\over \pi\mu_1 b^2}{\partial A\bar{\partial}B
\over (A+B)^2}\right)\, ,
\ee

\be\label{phi2}
\phi_2={1\over 2b}\log\left({1\over \pi\mu_2 b^2}{\partial C\bar{\partial}B
\over (C+B)^2}\right)\, ,
\ee
where
\be\label{cphik}
C={\zeta A+\beta\over \gamma A+\delta}\, .
\ee
Substituting (\ref{phi1}) and (\ref{phi2}) in (\ref{def3})
we find that it is satisfied with
\be\label{lambdik}
e^{-\Lambda b}={A-C\over \sqrt{\partial A\partial C}}\, .
\ee
Since $A$ and $C$ are holomorphic functions, $\Lambda$
is holomorphic as well, as it is stated in (\ref{hollam}).

It is straightforward to check that (\ref{def111}) is satisfied
as well with $\phi_1$, $\phi_2$ and $\Lambda$
given by
 (\ref{phi1}), (\ref{phi2}) and (\ref{lambdik})
respectively.
And finally inserting (\ref{phi1}), (\ref{phi2}) and (\ref{lambdik})
in (\ref{kapik})  we see that it is also fulfilled with
\be\label{aldet}
\kappa={\zeta+\delta\over 2}\, .
\ee

\section{$X_s$ defects in the heavy asymptotic limit}

In this section we link the continuous family of defects constructed in section 3 with the Lagrangian constructed in the previous section.
For this purpose we will use the heavy semiclassical asymptotic limit. Recall that in all semiclassical limits, one takes $b\to 0$ and
the action blows to infinity like $b^{-2}$.
It is known \cite{Zamolodchikov:1995aa} that in the heavy asymptotic limit, when one additionally scales  $\alpha={\eta\over b}$ and
$\Delta_{\alpha}=\eta(1-\eta)/b^2$, correlation functions are given by the exponential of the regularized action with the inserted fields computed on solution
with the logarithmic singularities around the insertion points.
The regularization is necessary to keep the action finite, since singularities of the solution may render it divergent.
So we should  compute the heavy asymptotic limit of the defect two-point function  (\ref{contt}) and compare
with the regularized defect action computed on the solutions with two singularities.

First we compute the heavy asymptotic limit of the defect two-point function  (\ref{contt}).
As we said before, in the heavy asymptotic limit we set   $\alpha={\eta\over b}$,
and also $s={\sigma\over b}$. Denote $\lambda_i=\pi\mu_i b^2$, $i=1,2$. We assume that $\eta$ is real and satisfies the Seiberg bound
$\eta< {1\over 2}$.
Performing the same steps  as in \cite{Poghosyan:2015oua} we can write in the heavy asymptotic limit
\footnote{Since we consider here only the case of real $\eta$ and real solutions of the Liouville equation,
we do not write here imaginary terms, which one has in  \cite{Poghosyan:2015oua}.}
\be
\langle V_{\alpha}(z_1,\bar{z}_1)X_sV_{\alpha}(z_2,\bar{z}_2)\rangle\sim\exp\left(-S^{\rm def}\right)\, ,
\ee
where
\bea\label{sdefect}
&&b^2S^{\rm def}=
4\eta(1-\eta)\log|z_1-z_2|
-\left({1\over 2}-\eta\right)\log\lambda_1-\left({1\over 2}-\eta\right)\log\lambda_2-\quad\\ \nonumber
&&(4\eta-2)\log(1-2\eta)+(4\eta-2)-2\pi|\sigma|(1-2\eta)\, .
\eea
Here we dropped all the terms in the exponential which blows slower then $b^{-2}$.
As it is explained in \cite{Poghosyan:2015oua} the required classical solution with two singular points can be built taking:
\be\label{aiki}
A(z)=e^{2\nu_1}(z-z_1)^{2\eta-1}(z-z_2)^{1-2\eta}\, .
\ee
\be\label{zaiki}
B(\bar{z})=-(\bar{z}-\bar{z}_1)^{1-2\eta}(\bar{z}-\bar{z}_2)^{2\eta-1}\, ,
\ee
\be\label{aiki0}
C(z)=e^{2\nu_2}(z-z_1)^{2\eta-1}(z-z_2)^{1-2\eta}=e^{2(\nu_2-\nu_1)}A(z)\, ,
\ee
Eq. (\ref{aldet}) implies
\be\label{coshik}
\kappa=\cosh(\nu_2-\nu_1)\, .
\ee
Inserting (\ref{aiki})-(\ref{aiki0}) in (\ref{phi1} and (\ref{phi2}) we obtain

\bea\label{solod11}
&&\varphi_1=-\log\lambda_1+2\log(1-2\eta)\\ \nonumber
&&-2\log\left(
{e^{\nu_1}|z-z_1|^{2\eta}|z-z_2|^{2-2\eta}\over |z_1-z_2|}-
{e^{-\nu_1}
|z-z_1|^{2-2\eta}|z-z_2|^{2\eta}\over |z_1-z_2|}\right)\, ,
\eea

\bea\label{solod22}
&&\varphi_2= -\log\lambda_2+2\log(1-2\eta)\\ \nonumber
&&-2\log\left(-
{e^{\nu_2}|z-z_1|^{2\eta}|z-z_2|^{2-2\eta}\over |z_1-z_2|}+
{e^{-\nu_2}
|z-z_1|^{2-2\eta}|z-z_2|^{2\eta}\over |z_1-z_2|}\right)\, .
\eea
Here $\varphi_i=2b\phi_i$, $i=1,2$.

The leading terms of $\varphi_1$ around $z_1$ are
\be
\varphi_1\to -4\eta\log|z-z_1|+X_1\, ,
\ee
where
\be\label{xx11}
X_1=-\log\lambda_1+2\log(1-2\eta)-(2-4\eta)\log|z_1-z_2|-2\nu_1\, .
\ee

The leading terms of $\varphi_2$ around $z_2$ similarly are
\be
\varphi_2\to -4\eta\log|z-z_2|+X_2\, ,
\ee
where
\be\label{xx22}
X_2=-\log\lambda_2+2\log(1-2\eta)-(2-4\eta)\log|z_1-z_2|+2\nu_2\, .
\ee

Since we consider here only insertions of the bulk field, and do not consider insertions of the defect or boundary fields, following the same steps as
in \cite{Zamolodchikov:1995aa} the regularized
action, with $n$ fields inserted  in the upper half-plane, and $m$ fields inserted in the lower half-plane, takes the form:
\bea\label{topdefbb}
&&b^2S^{\rm top-def}={1\over 8\pi i}\int_{\Sigma_1^R-\cup_i d_i}\left(\partial\varphi_1\bar{\partial} \varphi_1+4\lambda_1 e^{\varphi_1}\right)d^2 z\\ \nonumber
&-&\sum_{i=1}^n\left({\eta_i\over 2\pi}\oint_{\partial d_i}\varphi_1 d\theta_i+2\eta_i^2\log \epsilon_i\right)
+{1\over 2\pi}\int_{{s_R}_1}\varphi_1 d\theta+\log R\\ \nonumber
&+&{1\over 8\pi i}\int_{\Sigma_2^R-\cup_j d_j}\left(\partial\varphi_2\bar{\partial} \varphi_2+4\lambda_2 e^{\varphi_2}\right)d^2 z\\ \nonumber
&-&\sum_{j=n+1}^{n+m}\left({\eta_j\over 2\pi}\oint_{\partial d_j}\varphi_2 d\theta_j+2\eta_j^2\log \epsilon_j\right)
+{1\over 2\pi}\int_{{s_R}_2}\varphi_2 d\theta+\log R\\ \nonumber
&+& \int_{-R}^R\left[-{1\over 8\pi}\varphi_2\partial_{\tau}\varphi_1+
{1\over 8\pi}\tilde{\Lambda}\partial_{\tau}(\varphi_1-\varphi_2)+{\sqrt{\lambda_1\lambda_2}\over 2\pi}
e^{(\varphi_1+\varphi_2-\tilde{\Lambda})/2}\right.\\ \nonumber
&&-{1\over \pi }
e^{\tilde{\Lambda}/2 }\left.\left({1\over 2}\sqrt{\lambda_1\over \lambda_2}e^{(\varphi_1-\varphi_2)/2}+{1\over 2}\sqrt{\lambda_2\over \lambda_1}e^{-(\varphi_1-\varphi_2)/2}-\kappa\right)\right]{d\tau\over i}\, .
\eea
where $\tilde{\Lambda}=2b\Lambda$, $\Sigma_i^R$ is a half-disc of the radius $R$ and  ${s_R}_i$ is a semicircle of the radius $R$ in the half-plane $\Sigma_i$, $i=1,2$.
The two-point function in question is given by the exponential of the regularized action with one field inserted in the upper half-plane, and
with one field inserted in the lower half-plane, with $\eta_1=\eta_2=\eta$, calculated on solution  (\ref{solod11}) and (\ref{solod22}).
To calculate it, we will use, that it satisfies the equation \cite{Zamolodchikov:1995aa,Poghosyan:2015oua,Harlow:2011ny}:
\be\label{sx1x2}
b^2{\partial S^{\rm top-def}_{\rm cl}\over \partial\eta}=-X_1-X_2\, .
\ee
Inserting (\ref{xx11}) and (\ref{xx22}) in (\ref{sx1x2}) one obtains
\be\label{sx12}
b^2{\partial S^{\rm top-def}_{\rm cl}\over \partial\eta}=
\log\lambda_1+\log\lambda_2-4\log(1-2\eta)+(4-8\eta)\log|z_1-z_2|+2(\nu_1-\nu_2)\, .
\ee

Integrating equation (\ref{sx12}) we obtain:
\bea\label{sdefecter}
&&b^2S^{\rm top-def}=
4\eta(1-\eta)\log|z_1-z_2|\\ \nonumber
&&+\eta\log\lambda_1+\eta\log\lambda_2-
(4\eta-2)\log(1-2\eta)+4\eta+2\eta(\nu_1-\nu_2)+C\, ,
\eea
where $C$ is a constant.

To fix the constant term we can directly compute the action (\ref{topdefbb})  for the solution  (\ref{solod11})-(\ref{solod22}) with
$\eta=0$:

\be\label{valk1}
\varphi_1=-\log\lambda_1-\log\left({e^{\nu_1}\over |z_1-z_2|}|z-z_2|^2-
{e^{-\nu_1}\over |z_1-z_2|}|z-z_1|^2\right)^2\, ,
\ee

\be\label{valk2}
\varphi_2=-\log\lambda_2-\log\left({e^{\nu_2}\over |z_1-z_2|}|z-z_2|^2-
{e^{-\nu_2}\over |z_1-z_2|}|z-z_1|^2\right)^2\, .
\ee

Evaluation of the action (\ref{topdefbb}) on the solution (\ref{valk1}), (\ref{valk2})
can be carried out along the same steps as done in appendix C of \cite{Poghosyan:2015oua}. The result is

\be\label{bs}
b^2S_0=-{1\over 2}\log \lambda_1-{1\over 2}\log \lambda_2-2-(\nu_1-\nu_2)\, .
\ee

Comparing (\ref{bs}) with (\ref{sdefecter}) fixes the constant $C$:
\be
C=-{1\over 2}\log \lambda_1-{1\over 2}\log \lambda_2-2-(\nu_1-\nu_2)\, .
\ee

Inserting this value of  $C$ in  (\ref{sdefecter}) we indeed obtain (\ref{sdefect})
if we set

\be
2\pi\sigma=\nu_1-\nu_2\, .
\ee

\end{document}